\documentclass[aps,prl,reprint]{revtex4-1}

\usepackage{graphicx}
\usepackage{dcolumn}
\usepackage{rotating}
\usepackage{adjustbox}
\usepackage[graphicx]{realboxes}
\usepackage{bm}
\usepackage{amsmath}
\usepackage{xcolor}

\begin{document}

\title{Constraining the Woods-Saxon Potential in Fusion Reactions Based on the Neural Network}

\author{Zepeng Gao$^{1}$}
\author{Siyu Liu$^{1}$}
\author{Peiwei Wen$^{2}$}
\author{Zehong Liao$^{1}$}
\author{Yu Yang$^{1}$}
\author{Jun Su$^{1,3}$}
\author{Yongjia Wang$^{4}$}
\author{Long Zhu$^{1,3,}$}

\email{Corresponding author: zhulong@mail.sysu.edu.cn}

\affiliation{
$^{1}$Sino-French Institute of Nuclear Engineering and Technology, Sun Yat-sen University, Zhuhai 519082, China\\
$^{2}$China Institute of Atomic Energy, Beijing 102413, China\\
$^{3}$Guangxi Key Laboratory of Nuclear Physics and Nuclear Technology, Guangxi Normal University, Guilin 541004, China\\
$^{4}$School of Science, Huzhou University, Huzhou 313000, China\\
}%

\date{\today}

\begin{abstract}

 The accurate determination of the nuclear interaction potential is essential for predicting the fusion cross sections and understanding the reaction mechanism, which plays an important role in the synthesis of superheavy elements. In this work, the neural network, which combines with the calculations of the fusion cross sections via the Hill-Wheeler formula, is developed to optimize the parameters of the Woods-Saxon potential by comparing the experimental values. The correlations between the parameters of Woods-Saxon potential and the reaction partners, which can be quantitatively fitted to a sigmoid-like function with the mass numbers, have been displayed manifestly for the first time. This study could promote the accurate estimation of nucleus-nucleus interaction potential in low energy heavy-ion collisions.

\end{abstract}

\maketitle

\section{Introduction}

In recent decades, the study of heavy-ion fusion reactions has garnered growing interest in the field of nuclear physics due to their importance for extending the period table of elements, as well as the understanding of the interplay between nuclear structure and the reaction dynamics \cite{canto2006fusion,canto2015recent,back2014recent,jiang2007expectations,karpov2017modeling}. However, the complexity of these reactions presents significant challenges for both experimental and theoretical investigations. One of the key challenges is accurately determining the nuclear interaction potential, which is crucial for investigating the reaction mechanism, but remains an arduous task due to several factors, including the quantum many-body problem, the form of nuclear force, among other complex influences \cite{umar2021pauli,mukherjee2007failure,bhuyan2020effect,alavi2017investigation}.

As one of most successful phenomenological forms of nuclear potential, the Woods-Saxon potential has been widely used to describe the nuclear interaction in heavy-ion fusion reactions, especially for the synthesis of the superheavy nuclei. However, the accuracy is limited by the uncertainties in the parameters. Much effort has been made to constrain the parameters in recent years \cite{wang2008fusion,jun2007modified,dudek1978new,zhang2022optimized}.
Nevertheless, due to a complex function with multiple and correlated parameters as well as existing computational limitations, the majority of studies primarily apply constraints on the parameters for specific systems, such as focusing solely on $^{12}\mathrm{C}$ or $^{16}\mathrm{O}$ induced reactions \cite{luo2022bayesian,evers2008systematic}. 
Therefore, it is imperative to accurately and efficiently constrain and optimize the Woods-Saxon potential parameters from a global perspective.


Machine learning, being adept at uncovering underlying patterns from vast amounts of data, as well as accurately fitting and predicting data, have garnered significant interest across various fields. 
 In the realm of nuclear physics, they hold immense potential for addressing challenges in terms of nuclear theories and experiments \cite{he2023machine,wang2023machine,carleo2019machine,boehnlein2022colloquium}. For example, machine learning methods have demonstrated successful applications in predicting nuclear masses \cite{gao2021machine,niu2018nuclear,wu2022multi,ming2022nuclear}, charge radii \cite{wu2020calculation,dong2022novel,shang2022prediction}, half-lives \cite{niu2019predictions,li2022deephalf,gao2023investigation}, nuclear energy density functional \cite{wu2022nuclear}, fission product yields \cite{wang2019bayesian,wang2022bayesian}, radionuclide diffusion \cite{feng2023application}, as well as facilitating studies on nuclear reactions \cite{li2021application,nijs2021bayesian,ma2020isotopic,song2022target,ma2022precise,wang2022decoding,li2023importance,wang2021modeling,wang2021finding,song2021determining,li2020application,he2021machine}. In recent years, although great progress has been made in applying machine learning methods to address issues in nuclear physics, it is worth to note that most of those studies are data-driven approaches. Data in a real process are governed by physical laws, thus, it is imperative to integrate the fundamental principles of physics into machine learning methods, enabling them to be guided by the laws of physics rather than relying solely on data-driven approaches. 

In this work, we draw inspiration from physics-informed neural network (PINN), aiming to incorporate physical information into the neural network. We will explore the capability of the neural network to constrain the parameters of the Woods-Saxon potential model by comparing the calculations of fusion cross sections using the Hill-Wheeler formula with the corresponding experimental values. Subsequently, we will explicitly demonstrate the correlations between the parameters and colliding partners. The outcomes of this investigation will be highly significant in enhancing our comprehension of heavy ion fusion reactions and will offer a fresh perspective for exploring nuclear physics conundrums.

\section{Models and Methods}
\subsection{Fusion cross section and Woods-Saxon potential}

At a given center-of-mass energy, the fusion cross section $\sigma_{\mathrm{fusion}}\left(E_{\mathrm{c} . \mathrm{m} .}\right)$ can be expressed as the sum of the cross section at each partial wave $J$:
\begin{equation}
\label{equation:1}
\sigma_{\mathrm{fusion}}\left(E_{\mathrm{c} . \mathrm{m} .}\right)=\frac{\pi \hbar^2}{2 \mu E_{\text {c.m. }}} \sum_{J=0}^{J_{\max }}(2 J+1) T\left(E_{\text {c.m. }}, J\right),
\end{equation}
where $\mu$ is the reduced mass, $T$ is the penetration probability and $J$ is the incident angular momentum. Note that we have specifically selected relatively light combinations of target and projectile, where the occurrence of quasi-fission is relatively minimal. The Hill-Wheeler formula \cite{hill1953nuclear}, by considering the efficient iterative capabilities of neural networks, proves to be a highly effective approximation for numerical solutions, especially in the fusion cross sections above the Coulomb barrier \cite{wang2017systematics}. It is a well-known analytical expression for the penetration probability at definite center-of-mass energy $E_{\rm c.m.}$ and angular momentum $J$:
\begin{equation}
\begin{aligned}
\label{equation:2}
& T_{\mathrm{HW}}\left(E_{\mathrm{c} . \mathrm{m} .}, J\right) \\
& \quad=\left\{1+\exp \left[\frac{2 \pi}{\hbar \omega(J)}\left(\frac{\hbar^2 J(J+1)}{2 \mu R_{\mathrm{B}}^2(J)}+B-E_{\mathrm{c} . \mathrm{m} .}\right)\right]\right\}^{-1},
\end{aligned}
\end{equation}
where $B$ denotes the barrier height for head-on collision. $R_{\mathrm{B}}(J)$ and ${\hbar \omega(J)}$ correspond to the position and curvature of the barrier under the $J^\mathrm{th}$ partial wave, respectively. The barrier curvature ${\hbar \omega(J)}$ can be calculated using the following formula:

\begin{equation}
\hbar \omega(J)=\left.\sqrt{-\frac{\hbar^2}{\mu} \frac{\partial^2}{\partial R^2} V(R, J)}\right|_{R=R_{\mathrm{B}}(J)}.
\end{equation}
The interaction potential $V(R,J)$ consists of a long-range Coulomb potential, a short-range nuclear potential, and a centrifugal potential, which can be expressed as follows:
\begin{equation}
V(R, J)=V_{\mathrm{C}}(R)+V_{\mathrm{N}}(R)+V_{\mathrm{R}}(R, J).
\end{equation}
The fusion cross sections at the sub-barrier, especially the deep sub-barrier region strongly depends on the coupling of the nuclear structures. We would like to emphasize that in order to minimize the uncertainties and focus on the parameters of Woods-Saxon potential, the fusion reactions occurring above the barrier are only investigated. The Coulomb potential and centrifugal potential can be expressed by the forms:
\begin{equation}
\begin{aligned}
 &V_{\mathrm{C}}\left(R\right)=\frac{Z_{\rm P} Z_{\rm T} e^2}{R},\\
 &V_{\mathrm{R}}\left(R, J\right)=\frac{\hbar^2 J(J+1)}{2 \mu R^2}.
\end{aligned}
\end{equation}
The Woods-Saxon nuclear potential can be expressed as \cite{singh2013analysis}:
\begin{equation}
\begin{aligned}
&V_{\mathrm{N}}\left(R\right)=\frac{-V_0}{1+\exp \left[\left(R-R_{\mathrm{P}}-R_{\mathrm{T}}\right) / a\right]},
\label{equation:V_N}
\end{aligned}
\end{equation} 
with
\begin{equation}
R_i = r_{0i}A^{1/3}_i, i = \rm{P, T},
\end{equation} 
where the depth $V_\mathrm{0}$, the radius parameter $r_\mathrm{0}$, and the diffuseness parameter $a$ are the major parameters of the Woods-Saxon potential. 
By changing the parameters of the Woods-Saxon potential, we can modify the position, height as well as curvature of the potential barrier, which ultimately affects the prediction of the fusion cross sections. Therefore, we can evaluate the optimization results of the parameters by comparing the predicted cross sections via neural networks with experimental values.
\begin{figure*}[hbtp]
\includegraphics[width=0.8\textwidth]{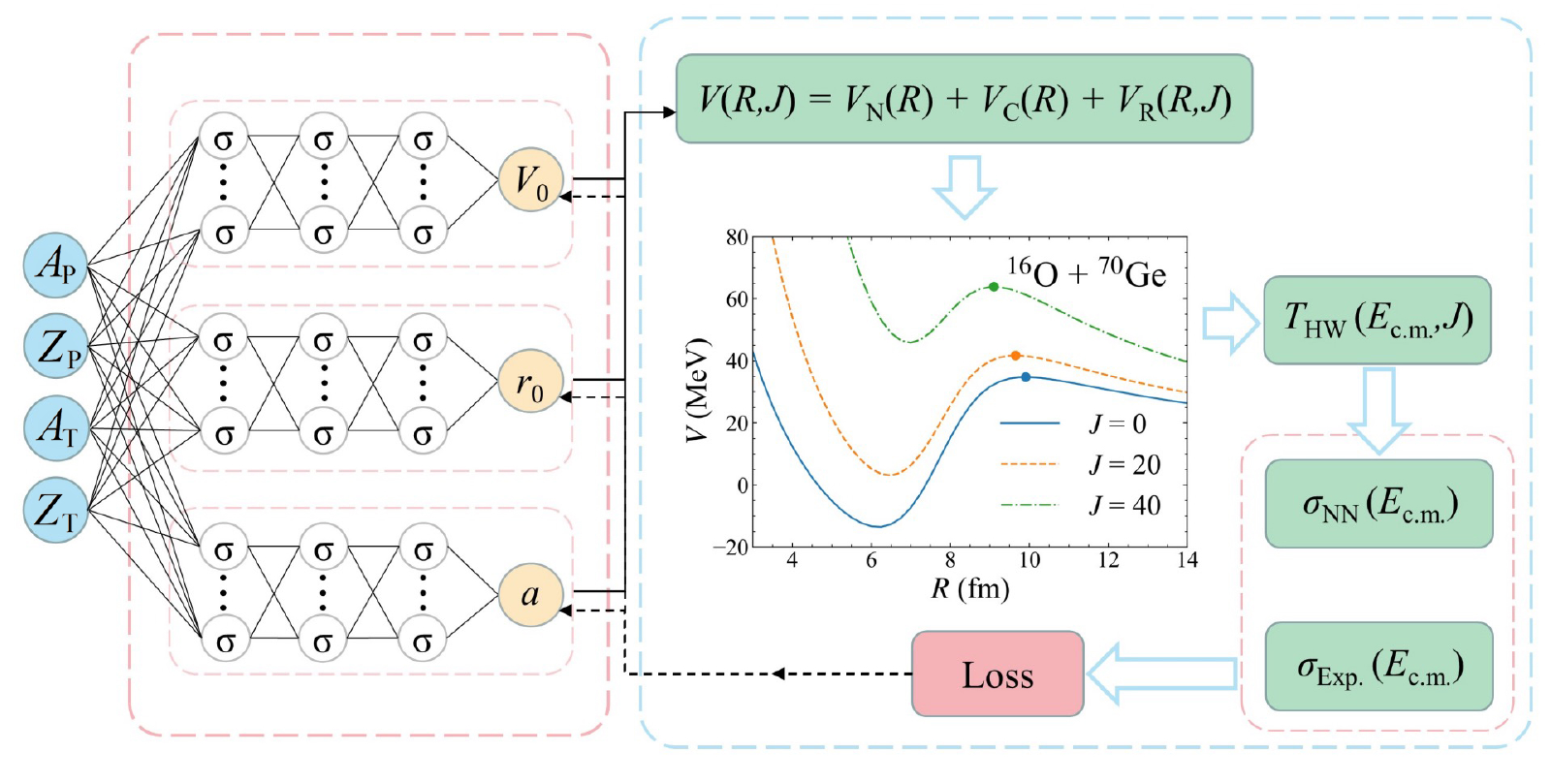}
\caption{\label{fig:PINN} The architecture of the neural network used in this work. It combines neural networks and process of the calculation of fusion cross-sections via Hill-Wheeler formula.}
\end{figure*}

\subsection{The neural network connected with the calculations of the fusion cross sections }


In this work, we perform calculations of fusion cross sections and train the neural network by comparing it with experimental data, as illustrated in Fig. \ref{fig:PINN}. Initially, $Z$ (proton number) and $A$ (mass number) of the projectile and target nuclei as the input features are fed into a neural network, which then generates three parameters corresponding to the Woods-Saxon potential. Moreover, the nuclear interaction potential at various angular momenta and the associated input parameters required for the Hill-Wheeler formula are obtained. By summing the capture probability over all angular momenta, the corresponding fusion cross sections can be derived. By comparing the fusion cross-section calculated using Hill-Wheeler formula with parameters $V_0$, $r_0$ and $a$ obtained from neural networks to the experimental values, we can establish a loss function, as expressed below:
\begin{equation}
\label{equation:8}
\mathcal{L}(i)=\left\{\sum_{e}\left[\lg\left(\sigma_{\mathrm{NN}}^{(i)}(e)\right)-\lg\left(\sigma_{\exp}^{(i)}(e)\right)\right]\right\}^2,
\end{equation}
where $e$ represents the $E_{\rm c.m.}$ associated with the experimental value in the $i$th projectile-target combination. It is important to note that for each set of parameters of Woods-Saxon potential, a fusion excitation function is calculated and compared with all experimental values associated with the projectile-target combination. Consequently, the resulting loss is not solely determined by each individual energy, but rather by considering the combined outcomes of all the $E_{\rm c.m.}$ with the specific system. Obtaining the loss allows for the training of the neural network, which will be discussed in the subsequent context.

Foremost, it is crucial to clarify that $\mathcal{L}(i)$ is not attributed to the parameters of Woods-Saxon potential, but rather to the fusion cross sections. Moreover, there is no explicit functional association between the fusion cross sections and the parameters of Woods-Saxon potential, so that it is not feasible to directly obtain the loss of three parameters using Backpropagation algorithm. However, the qualitative association can be analyzed and the existence of a positive correlation between them has been established in advance. Therefore, we can attribute the loss of the parameters of Woods-Saxon potential to the loss of fusion cross sections and assign suitable weight coefficients in a straightforward manner, as expressed below:
\begin{equation}
\mathcal{L}(i)^{(j)}=\alpha^{(j)}\mathcal{L}(i), j = V_0, r_0, a.
\end{equation}
The weight coefficients, by taking into account the empirical ranges, were assigned as 1, 0.001, and 0.01 for $V_0$, $r_0$ and $a$, respectively. It is worth to emphasize that the weight coefficients weakly influence the behavior of correlations found in this work. After obtaining the loss of the parameters, the Backpropagation algorithm can be employed for training the neural network.

The multi-output neural network consists of three neural networks, wherein each network comprises three hidden layers with 32, 64, and 128 neurons, respectively. ``Tanh” and ``Adam” were selected as the activation function and optimizer. The learning rate $\alpha$ started at $10^{-4}$ and gradually decreases to $10^{-6}$ during the training process. Furthermore, a total of 343 reactions from the online dataset \cite{dataset} were chosen for this study, where $Z_\mathrm{P} Z_\mathrm{T} \leq 1600$ and there were at least three experimental data points above the Coulomb barrier. To select the data, the Coulomb barrier can be estimated by the latest fitting formula \cite{wen2022new} as:
\begin{equation}
V_{\mathrm{B}}=\frac{Z_{\mathrm{P}} Z_{\mathrm{T}} e^2}{0.9782\left(A_{\mathrm{P}}^{1 / 3}+A_{\mathrm{T}}^{1 / 3}\right)+4.2833}.
\end{equation}
30 projectile-target combinations of the dataset were randomly selected as the testing set and validation set, respectively. Thus, the remaining 283 combinations were utilized as the training set and batch size was set to 70 during training.

\begin{figure}[h!]
\includegraphics[width=0.5\textwidth]{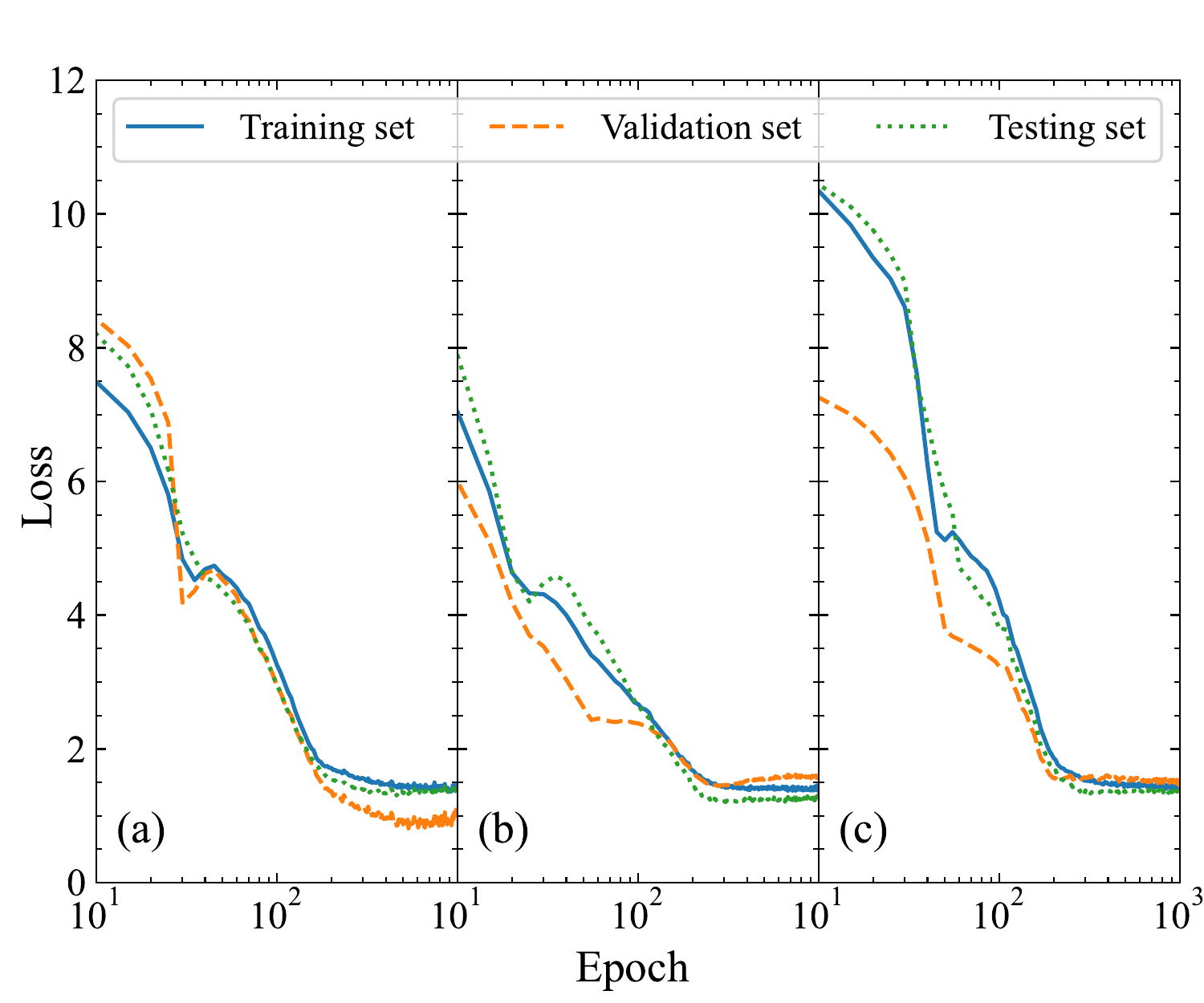}
\caption{\label{fig:LOSS} The loss curves of three combinations of the training and validation sets as functions of the epoch (training time). Solid, dash and dotted line denote the training, validation, and testing sets, respectively.}
\end{figure}

\begin{figure*}[htp]
\includegraphics[width=1\textwidth]{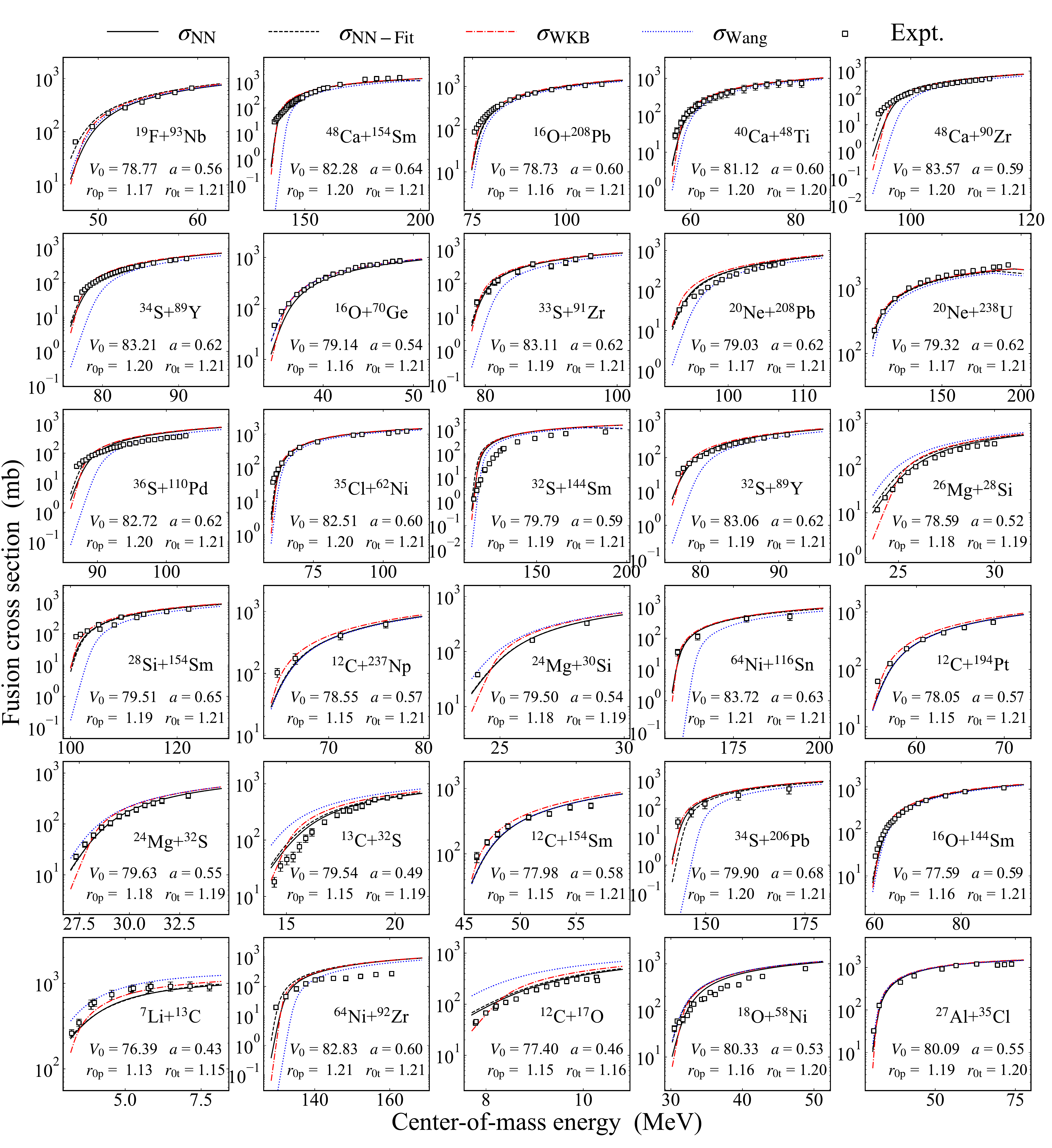}%
\caption{\label{fig:testing pred} The experimental and predicted fusion excitation functions for testing set. The fusion cross sections obtained from neural network and other theoretical models explained in Tab. \ref{tab:1} are shown with different linetypes, while the squares denote experimental data. The parameters constrained by neural network are also displayed.}
\end{figure*}

The loss curves of training set, validation set and testing set, by training on 283 combinations through multiple optimizations and structural adjustments, are displayed in Fig. \ref{fig:LOSS}. The loss function can be defined as the average value of the individual loss function $\mathcal{L}(i)$ computed for each projectile-target combination:
\begin{equation}
\label{equation:11}
   {\rm Loss} = \frac{1}{N}\sum_{i}^{N}\mathcal{L}(i).
\end{equation}
The loss for the validation and testing sets initially decreases and then plateaus or slightly increases, whereas the loss for the training set continuously decreases with the increase of epoch. Hence, a truncation, to prevent overfitting and to obtain the extrapolation ability in the neural network, was employed at epoch = 500. Furthermore, one can find that the loss for the training set is higher than that for the validation set, as shown in Fig. \ref{fig:LOSS}(a). It does not necessarily imply underfitting, since the loss on the validation set has already plateaued. We further trained two neural networks using random combinations of training and validation sets, and the corresponding loss curves are shown in Fig. \ref{fig:LOSS}(b)(c), where the loss for the training set is lower than that for the validation set. This discrepancy is mainly due to the impact of data sampling on the training and validation sets, since the fusion cross-sections of certain systems are challenging to precisely express via the Hill-Wheeler formula. However, this effect is not significant for subsequent correlation analysis.

\section{Results and discussions}

The extrapolation ability of the neural networks can be demonstrated by its predictions on fusion cross sections in the testing set, which has not been through the training process. 
The neural network predictions thus have been compared with the experimental data, as shown in Fig. \ref{fig:testing pred}.

One can see that the extrapolation capability of the neural network is reliable, as the predicted cross section values ($\sigma_{\rm NN}$) are in good agreement with the experimental data for most of the reaction systems. However, in light projectile-target combination, there are some discrepancies between neural network predictions and experimental data. This is primarily due to the strong structural effects that exist in such systems, which can lead to the possibility of cluster formation.

Furthermore, we employed two additional cross section calculation methods to evaluate the neural network's extrapolation ability and its constraint on parameters of Woods-Saxon potential, which are the $\sigma_{\rm WKB}$ and the $\sigma_{\rm Wang}$ expounded in Tab. \ref{tab:1}. The Wentzel-Kramers-Brillouin (WKB) approximation \cite{miller1953wkb} is a classical method for calculating tunneling probability, which is widely used in studies on fusion or alpha decay. In addition, the $\sigma_{\rm Wang}$ is also used for evaluation with the same calculation of the tunneling probability using Eq. (\ref{equation:2}) but with different parameters obtained from Ref. \cite{wang2017systematics} with $V_0$ = 80 MeV, $r_0$ = 1.16 fm,  $a = \left\{1.17\left[1+0.53\left(A_\mathrm{P}^{-1/3}+A_\mathrm{T}^{-1/3}\right)\right]\right\}^{-1}\mathrm{~fm}$.

\begin{figure*}[htp]
\includegraphics[width=0.9\textwidth]{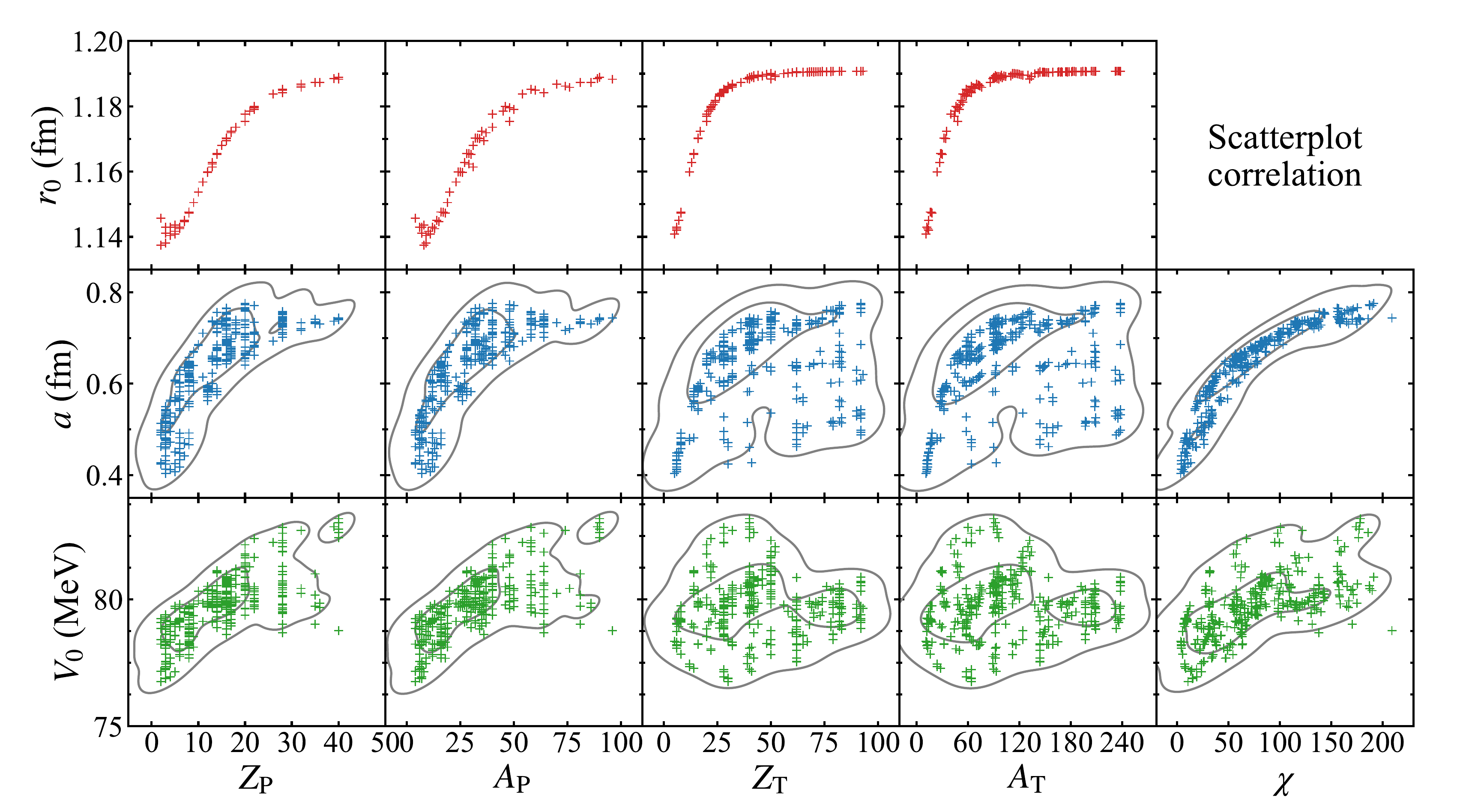}
\caption{\label{fig:scatter} Scatter plots display the correlation
between the horizontal and vertical labeled quantities. Each scatter point denotes one projectile-target combination.}
\end{figure*}

\begin{table}[htb]
\caption{
Calculation methods of fusion cross sections. Model and Parameter denote the theoretical model for calculating tunneling probability and the sources of Woods-Saxon potential parameters. Loss denotes the deviations between the model's results and the experimental values in the testing set, which are obtained from Eq. (\ref{equation:8}) and Eq. (\ref{equation:11}).}
\label{tab:1}
\begin{tabular*}{8cm} {@{\extracolsep{\fill} } lllc}
\toprule
 & Model & Parameter & Loss \\
\hline
    $\sigma_{\rm NN}$      & Hill-Wheeler & neural network & 1.26 \\
    $\sigma_{\rm NN-Fit}$  & Hill-Wheeler & Eq. (\ref{equation:14}), (\ref{equation:15}), (\ref{equation:16}) & 1.38 \\
    $\sigma_{\rm WKB}$     & WKB approximation & neural network & 1.33 \\
    $\sigma_{\rm Wang}$    & Hill-Wheeler & Ref. \cite{wang2017systematics} & 4.47 \\
\hline
\end{tabular*}
\end{table}
The fusion cross sections obtained from these two additional models on the testing set are also shown in Fig. \ref{fig:testing pred} along with the corresponding losses detailed in Tab. \ref{tab:1}. By comparing $\sigma_{\rm NN}$ and $\sigma_{\rm Wang}$, it is evident that without the constraints imposed by the neural network and relying solely on fixed parameter forms, the model fails to accurately replicate the experimental values, leading to a substantial increase in loss. This directly highlights the constraining capability of the neural network on the parameters. Additionally, when we incorporate the learned potential from neural network into the WKB calculation, the results ($\sigma_{\rm WKB}$) that closely resemble those obtained from the Hill-Wheeler formula are obtained. This observation signifies the neural network's robust generalization ability. The ability to extrapolate and generalize exhibited by the neural network have been confirmed, we can thus explore the correlation between potential parameters and the reaction partners.

Our aim is to explore the underlying correlations of the parameters of Woods-Saxon potential with the reaction systems. We show the parameters of Woods-Saxon potential constrained by the neural network in Fig. \ref{fig:scatter}.
The scatter plots for the parameters of mass number, proton number as well as Coulomb parameter ($\chi=Z_{\rm P}Z_{\rm T}/(A_{\rm P}^{1/3}+A_{\rm T}^{1/3})$) are shown.  The strong correlations indicate that the Woods-Saxon potential parameters could be constrained and characterized by these quantities. It can be seen that all of these parameters fall within a reasonable range. The radius parameter $r_0$ displays a rapid increase from 1.14 fm to approximately 1.18 fm, followed by a slower convergence to around 1.19 fm as the mass number $A$ and proton number $Z$ increase. Additionally, the depth parameter $V_0$ exhibits a discernable trend of increasing with increasing $A$, $Z$, and $\chi$, with values falling within the range of roughly 77-83 MeV. The surface diffuseness parameter $a$ shows an even more pronounced correlation, with a rapid increase followed by a slower rise as $A$, $Z$, and $\chi$ increase, within a range of approximately 0.4-0.8 fm. Note that the distributions of $V_0$ and $a$ exhibit a certain degree of diffuseness, primarily because these two parameters are shared by the entire combination and are not solely determined by either the projectile or the target nucleus. The Coulomb parameter is a measure of the direct correlation between the projectile-target combination, and there exists a strong linear correlation with the Coulomb barrier \cite{wen2022new,zhu2023law}, which directly affects the fusion cross sections and makes it highly relevant to the constrained $V_0$ and $a$. Indeed, there is not a single unique set of Woods-Saxon parameters capable of fully describing the experimental values. The strength of neural network lies in its capacity to identify more direct correlations between the reaction partners and the intricate parameter space.

\begin{figure}[h]%
    \centering
    \includegraphics[width=.4\textwidth,height=.3\textwidth]{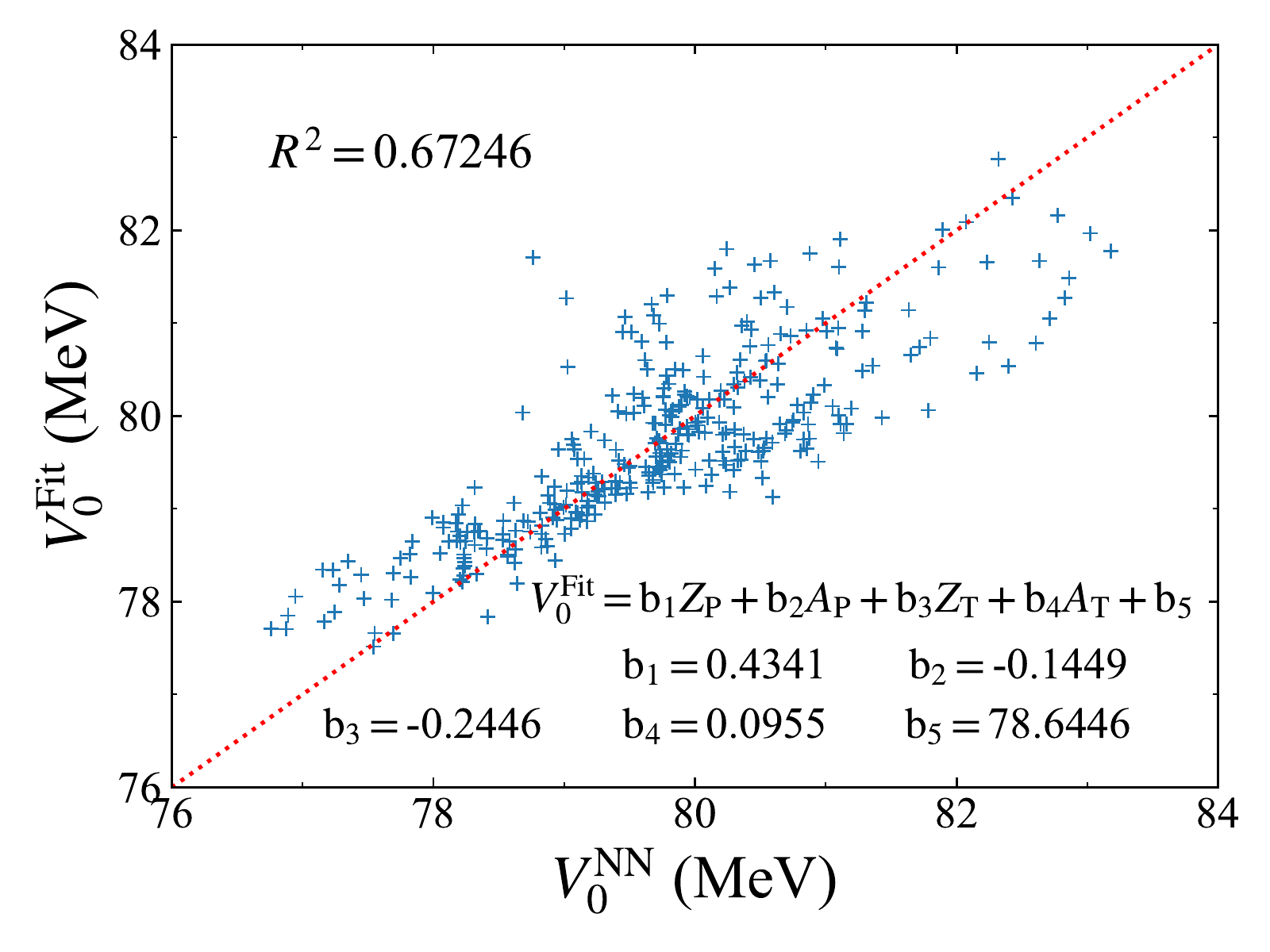}
    \includegraphics[width=.4\textwidth,height=.3\textwidth]{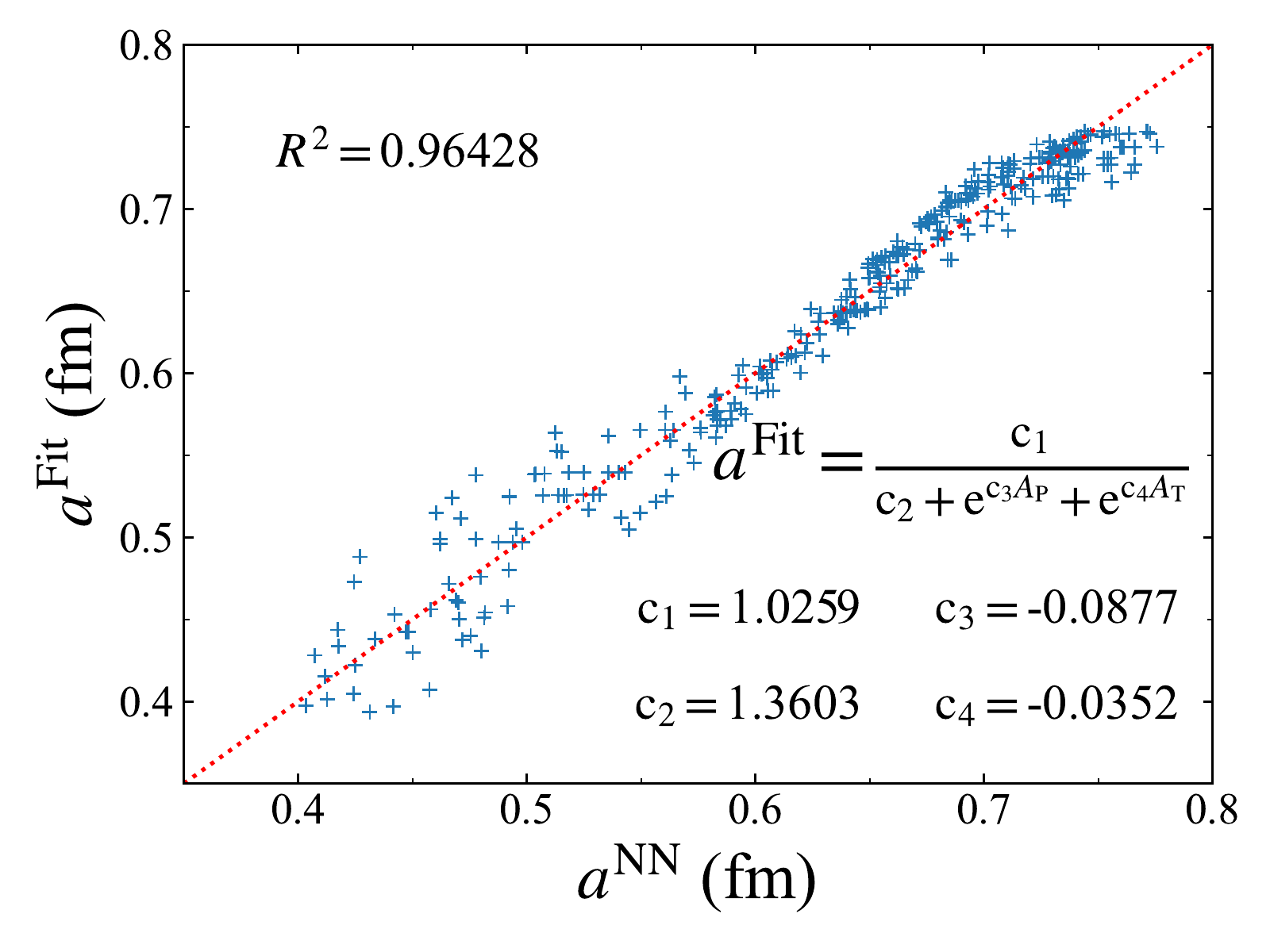}
    \includegraphics[width=.4\textwidth,height=.3\textwidth]{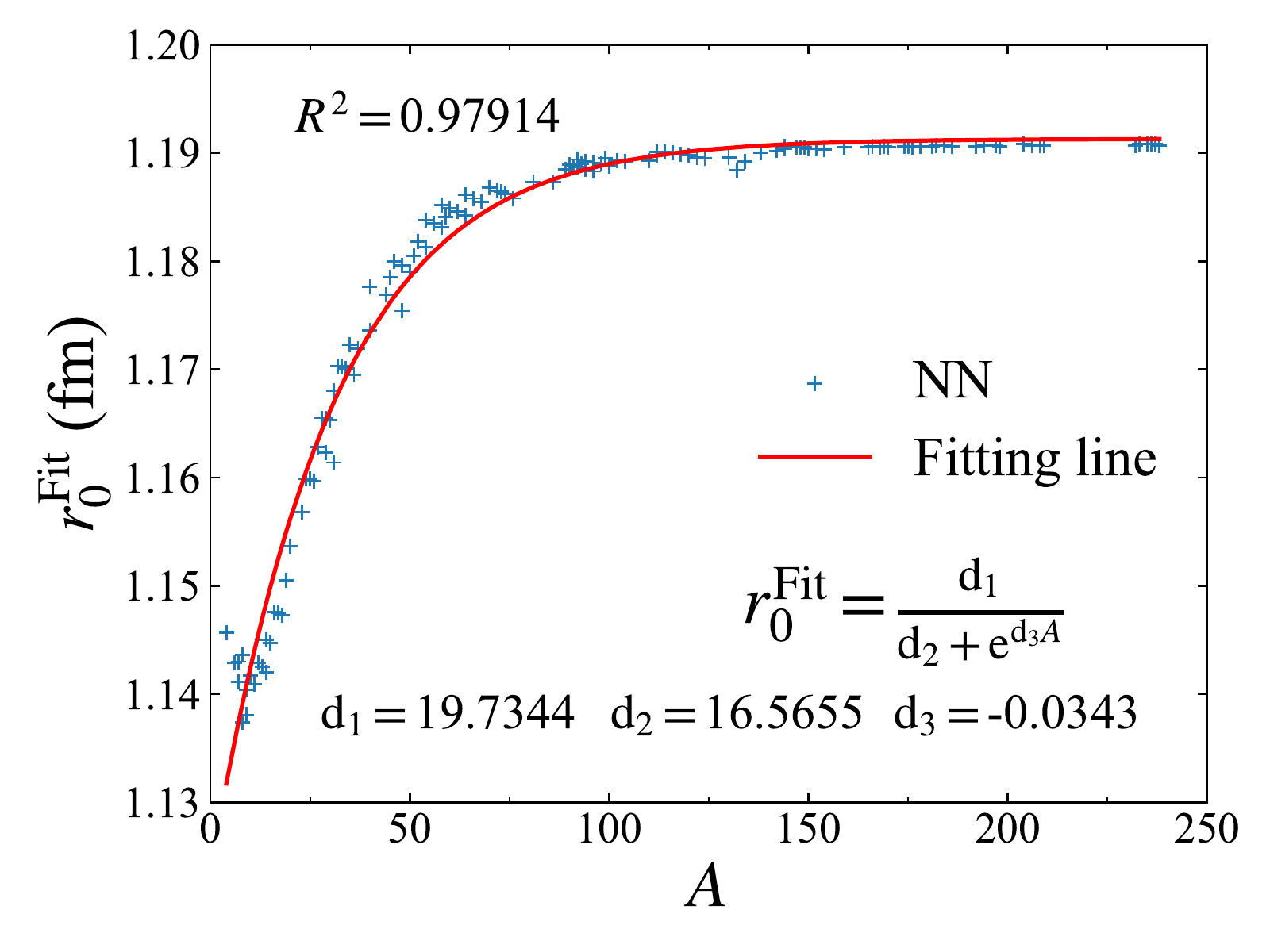}
    
\caption{Upper (Middle) panel: the predicted $V_0$ ($a$) from neural network versus the fitted $V_0$ ($a$). Each scatter point denotes one projectile-target combination. Lower panel: the predicted $r_0$ from neural network (dots) and the fitted $r_0$ (solid line) as the functions of mass number. Each scatter point denotes a projectile or target nucleus. The fitting formula and corresponding coefficients, as well as the coefficient of determination, are also displayed.}
\label{fig:fit}
\end{figure}

 The above trends and ranges between the three parameters of the Woods-Saxon potential and $A$, $Z$, and $\chi$ are considered reasonably in physics \cite{wang2008fusion,evers2008systematic,gautam2022comprehensive}. The Woods-Saxon potential describes the nuclear force potential in the nucleus-nucleus interaction, and the nuclear force is a short-range strong interaction attraction. Among them, $V_0$ controls the strength of the nuclear force, which increases with $A$ in the region of light partner. However, the correlation between $V_0$ and heavy collision partner is weak. This is because for large $A$, the value of potential is approximately flat in the center and the variation is weak.
  Currently, the available experimental data are mostly in the  reactions with stable projectiles and targets. Due to the curvature of the $\beta$ stable line, the spread of neutron skin is shown with increasing isospin. Consequently, the trend that the parameters $a$ and $r_0$ increase with increasing mass and charge number for both light and heavy partners can be seen. 

Based on the qualitative constraints mentioned above, we expected to obtain a more specific, intuitive, and analytical quantitative expression. As a result, the strongly correlated parameters $a$ and $r_0$ were fitted with the mass numbers of projectile-target combinations using a sigmoid-like function, while the relatively weak correlation of $V_0$ was also fitted with the proton numbers and the mass numbers into a linear function, as shown in Fig. \ref{fig:fit}. One can see that the weak correlation between $V_0$ and the reaction partners, as shown in Fig. \ref{fig:scatter}, leads to insufficient fitting accuracy. However, the good performance of fitting using a sigmoid-like function can be easily observed in middle panel of Fig. \ref{fig:fit} where the predicted $a$ is plotted versus the fitting results. The fitting results are better in regions where $a$ is larger, which is consistent with the broadening of $a$ observed in Fig. \ref{fig:scatter}. The fitting accuracy for $r_0$ is higher in lower panel of Fig. \ref{fig:fit}, mainly due to $r_0$ depends only on the individual projectile or target nucleus instead of the combination. The figure displays the fitted parameters and coefficient of determination ($R^2$), which provide an indication of the goodness-of-fit. The $R^2$ for $a$ and $r_0$ are greater than 0.96, which suggests that the fitting is highly effective. The final quantitative constraints on those parameters can be expressed as follows:

\begin{equation}
\label{equation:14}
V_{0}=\mathrm{b_{1}}Z_{\mathrm{P}}+\mathrm{b_{2}}A_{\mathrm{P}}+\mathrm{b_{3}}Z_{\mathrm{T}}+\mathrm{b_{4}}A_{\mathrm{T}}+\mathrm{b_{5}},
\end{equation}
where $\mathrm{b}_{1}=0.4341$ MeV, $\mathrm{b}_{2}=-0.1449$ MeV, $\mathrm{b}_{3}=-0.2446$ MeV, $\mathrm{b}_{4}=0.0955$ MeV and $\mathrm{b}_{5}=78.6446$ MeV.

\begin{equation}
\label{equation:15}
\mathit{a}=\frac{\mathrm{c}_{1}}{\mathrm{c}_{2}+\mathrm{e}^{\mathrm{c}_{3}{A_{\rm P}}}+\mathrm{e}^{\mathrm{c}_{4}{A_{\rm T}}}},
\end{equation}
where $\mathrm{c}_{1}=1.0259$ fm, $\mathrm{c}_{2}=1.3603$, $\mathrm{c}_{3}=-0.0877$, and $\mathrm{c}_{4}=-0.0352$.
\begin{equation}
\label{equation:16}
\mathit{r}_{\mathrm{0}i}=\frac{\mathrm{d}_{1}}{\mathrm{d}_{2}+\mathrm{e}^{\mathrm{d}_{3}{A_{i}}}}, i={\rm P, T},
\end{equation}
where $\mathrm{d}_{1}=19.7344$ fm, $\mathrm{d}_{2}=16.5655$, and $\mathrm{d}_{3}=-0.0343$.
$Z_{\rm P(T)}$ and $A_{\rm P(T)}$ denote the proton numbers and mass numbers of light (heavy) partner, respectively.

Similarly, we incorporated the fitted parameters mentioned above into the Woods-Saxon potential for cross-section calculations via Hill-Wheeler formula. The performance on the testing set is presented as $\sigma_{\rm NN-Fit}$ in Fig. \ref{fig:testing pred}. One can observe that the curves of $\sigma_{\rm NN}$ and $\sigma_{\rm NN-Fit}$ overlap almost completely for most reaction partners, and the loss of $\sigma_{\rm NN-Fit}$ only increases slightly, despite the insufficient fitting of $V_0$.

\section{Conclusions}

We have successfully incorporated the calculations of the fusion cross sections via the Hill-Wheeler formula into neural network and constructed the neural network framework to optimize the parameters of the Woods-Saxon potential in a more accurate and efficient way. Upon comparison with experimental fusion cross sections, the neural network demonstrated impressive predictive performance on the testing set, and effectively constrained the three parameters of the Woods-Saxon potential. The  extrapolation and generalization abilities of neural network are demonstrated by comparing the fusion cross-section calculated using the WKB method and other potential parameter forms. Furthermore, the constrained values of $a$ and $r_0$ exhibit a strong correlation with the projectile-target combinations, which are accurately fitted by a sigmoid-like function for the first time. Those functions can be conveniently used to model heavy-ion fusion reactions. 


\section{Acknowledgments}
This work was supported by the National Natural Science Foundation of China under
Grant No. 12075327; Fundamental Research Funds for the Central Universities, Sun Yat-sen University under Grant No. 23lgbj003; The Open Project of Guangxi Key Laboratory of Nuclear Physics and Nuclear Technology under Grant No. NLK2022-01; Guangdong Major Project of Basic and Applied Basic Research under Grant No. 2021B0301030006; Central Government Guidance Funds for Local Scientific and Technological Development, China (No. Guike ZY22096024); Director's Foundation of  Department of Nuclear Physics, China Institute of Atomic Energy (12SZJJ-202305)

\end{document}